\documentclass[aps,pra,twocolumn ,groupedaddress]{revtex4}

\usepackage{amsmath}
\usepackage{amssymb}
\usepackage{physics}
\usepackage{caption}
\usepackage{graphicx}
\usepackage[colorlinks = true, linkcolor = blue, urlcolor  = blue, citecolor = blue, anchorcolor = blue]{hyperref}
\usepackage{subfig}
\usepackage{comment}
\usepackage{multirow}
\usepackage[font=small,labelfont=bf,justification=justified,format=plain]{caption}
\usepackage{xcolor}
\usepackage{tabularx}
\usepackage{mathrsfs}

\newcommand{\chavg}{\langle C_{h}  \rangle}

\newcommand{\unit}{1\!\!1}
\begin{document}


\title{Restrictions to realize multiport quantum dense coding in a many-body quantum spin system with two- and three-body interactions}


\author{P. Kiran}
\affiliation{Department of Physics, Indian Institute of Technology Dharwad, Dharwad, Karnataka, India - 580011}

\author{Hemant Shreepad Hegde}
\affiliation{Department of Physics, Indian Institute of Technology Dharwad, Dharwad, Karnataka, India - 580011}

\author{Harsha Miriam Reji}
\affiliation{Department of Physics, Indian Institute of Technology Dharwad, Dharwad, Karnataka, India - 580011}

\author{R. Prabhu}
\affiliation{Department of Physics, Indian Institute of Technology Dharwad, Dharwad, Karnataka, India - 580011}



\begin{abstract}
   Quantum information with many-body quantum spin systems has, from
   time to time, given intriguing and intuitive outcomes to our
   understanding of multiport quantum communications. We identify that
   in an anisotropic many-body quantum spin system with two- and
   three-body interactions, when its two-spin subsystems are all
   negative under partial transpose, one can restrict this system for realizing only the multiport
   quantum dense coding protocol which has ($N-1$) senders and a single
   receiver. All other single and multi channel dense coding protocols
   will have quantum dense coding capacities less than that of their
   respective classical capacities. We characterize the multiport
   quantum dense coding capacity with ($N-1$) senders and a single
   receiver for this system with respect to its system parameters. We
   also define a {\em magnetic field averaged dense coding capacity} for
   this system, which allows us to comprehensively capture the influence
   of the entire range of external applied magnetic field and
   characterize its variation with respect to other system parameters.
\end{abstract}


\maketitle


\section{Introduction}\label{sec:introduction}
Quantum entanglement~\cite{horodecki2009quantum}, the characteristic
trait of quantum physics, forms the foundation for revolutionary
advancements in various quantum communication protocols, which enable the
development of quantum communication technologies that promise
unparalleled security and enhanced communication capabilities. Examples of
such quantum information protocols include quantum dense
coding~\cite{bennett1992communication}, quantum
teleportation~\cite{bennett1993teleporting}, quantum key
distribution~\cite{bennett2014quantum}, etc. Although there has been
a considerable focus on quantum communication protocols between two
parties, with extensive exploration on both the
experimental~\cite{gisin2002quantum, pan2012multiphoton,
haffner2008quantum} and theoretical~\cite{horodecki2009quantum,
bennett1992communication, bennett1993teleporting} fronts, there is a
compelling necessity to explore multiparty communication involving
multiple senders and receivers, as this exploration is crucial for the
actualization of quantum communication networks.

Quantum dense coding~\cite{bennett1992communication} is one of the
quantum communication protocols, which is easily implemented on
multiparty quantum systems, where these systems function as multiport
communication channels. This protocol facilitates the transmission of
classical information between two distant locations and improves the
capacity over its classical counterparts, with the help of an entangled
state which acts as a single channel shared between the sender and the receiver. The
quantum dense coding protocol has been demonstrated on various types of
physical systems, such as atomic qubits~\cite{schaetz2004quantum},
nuclear magnetic resonance~\cite{fang2000experimental}, displaced photon
states~\cite{podoshvedov2009displaced}, quantum spin
systems~\cite{prabhu2011disorder}, trapped ions~\cite{cai2015quantum} as
well as continuous variable states like Gaussian
states~\cite{lee2014continuous} and multimode squeezed states of
light~\cite{patra2022quantum}, etc. Initially designed for
point-to-point communication~\cite{aspect1981experimental,
tittel1998violation, gisin2007quantum, ursin2007entanglement}, the
quantum dense coding protocol has since evolved to accommodate
multiparty networks involving multiple senders and multiple
receivers~\cite{bose2003quantum, bruss2004distributed, das2014multipartite,
das2015distributed, han2023simultaneous}. 

Quantum spin systems have been studied from the perspective of various
quantum communication protocols~\cite{bose2003quantum, bayat2005thermal,
avellino2006quantum, cai2006decoherence, deng2008quantum,
burrell2009information, xiang2010quantum}, including quantum dense
coding protocol~\cite{fang2000experimental, prabhu2011disorder}. The
study of quantum communication related aspects using quantum spin
systems is usually restricted to Hamiltonians with only two-body
interactions~\cite{wang2012quantum, rong2012quantum, giorda2010gaussian,
adesso2010quantum, bellomo2012dynamics, prabhu2011disorder}. However,
quantum many-body systems may possess three- or higher-body interactions
along with the two-body ones. The theoretical analysis of three-body
interactions in quantum spin systems~\cite{titvinidze2003phase,
pachos2004effective, sende2010channel, bera2012multisite, liu2012chiral,
tsomokos2008chiral, peng2010ground-state, you2016quantum,
lou2004quantum, lou2005quantum, dealcantarabonfim2014quantum,
derzhko2011exact, shi2009effects, peng2009quantum,
zhang2010entanglement} and the experimental evidence of the existence of
three-body interactions in ultracold atoms~\cite{mark2011precision},
Rydberg atoms~\cite{lesanovsky2011many}, cold polar
molecules~\cite{buchler2007three, capogrosso2009phase} and various
physical substrates~\cite{pachos2004three, pachos2004effective,
feng2020quantum, liu2020synthesizing} offer valuable insights. These
findings lead us towards possible new research avenues in the
utilization of multiparty quantum spin systems featuring two- and
three-body interactions as multiport channels of quantum communication
protocols~\cite{aili2019dense, xi2015entanglement}.  Therefore, here we
focus our study on such a quantum spin system containing two- and
three-body interactions, which function as a multiport quantum channel
for realizing multiport quantum dense coding protocols with various
combinations of senders and receivers within this system.

The many-body quantum spin system considered here consists of two types
of two-body interactions, viz., XX and YY, with anisotropy existing
between these two types of interactions, the three-body interaction of
the form XZY$-$YZX, and an external applied magnetic field in the
Z-direction.  We study the multiport quantum dense coding capacity of
various channels formed by the ground state as well as its subsystems
when the Hamiltonian contains finite number of spins. In
Sec.~\ref{sec:hamiltonian}, the Hamiltonian of this anisotropic
multiparty quantum spin system with two- and three-body interactions
under the influence of an external magnetic field is described. In
Sec.~\ref{sec:qdc}, the quantum dense coding protocol between two
parties forming a single channel is succinctly presented along with the
extended definitions of quantum dense coding capacities for various
possible multi channel scenarios. In Sec.~\ref{sec:dcc_isotropic}, the
many-body quantum spin system considered here is studied for its quantum dense
coding capacities in the single and multi channel scenarios and the
cases which have the quantum dense coding advantage are recognized. The
multiport quantum dense coding capacity, which has the quantum dense
coding advantage, is then characterized with respect to the various
system parameters of the system under consideration to identify their
ranges for which maximum dense coding capacity is achieved. In order to capture
the behaviour of the multiport quantum dense coding capacity in the
entire range of external applied magnetic field strengths across the
system, we introduce a quantity called {\em magnetic field averaged
dense coding capacity} and study its variations with respect to other
system parameters. A brief conclusion is presented in
Sec.~\ref{sec:conclusion}.

\section{The Hamiltonian}\label{sec:hamiltonian}

Quantum spin systems
have been investigated for their potential application in various quantum
communication processes, such as quantum dense
coding~\cite{prabhu2011disorder, prabhu2013exclusion,
das2014multipartite, das2015distributed}, quantum state
transfer~\cite{bose2003quantum, avellino2006quantum, deng2008quantum},
quantum key distribution~\cite{hong2007quantum, bera2016information},
etc. Such studies primarily focus
on quantum spin systems with only two-body
interactions~\cite{sachdev2011quantum, wang2012quantum,
rong2012quantum, giorda2010gaussian, adesso2010quantum,
bellomo2012dynamics}, and has been extended to quantum spin systems with
higher-body interactions~\cite{guo2011pairwise, fu2017effect, titvinidze2003phase,
derzhko2011exact, lou2004quantum,
guo2011entanglement, kiran2023shareability}. Complex communication networks that use multiparty quantum spin
systems as multiport channels, similar to classical internet networks,
have been proposed recently~\cite{chepuri2023complex}. It has also been demonstrated that
quantum communication can be enhanced if quantum spin systems with
XZY$-$YZX type of three-body interaction along with Heisenberg XXZ type
of two-body interactions are considered~\cite{xiang2010quantum}. 

The Hamiltonian of one such anisotropic multiparty quantum spin system with two- and
three-body interactions, which is of our interest, is given by
\begin{eqnarray}
    H  &=& -\frac{J}{4} \sum_{n=1}^{N} \Big[ (1 + \gamma) \sigma_{n}^{x}
    \sigma_{n+1}^{x} + (1 - \gamma) \sigma_{n}^{y} \sigma_{n+1}^{y} \nonumber \\ 
    & & + \frac{\alpha}{2} \qty(\sigma_{n-1}^{x} \sigma_{n}^{z}
    \sigma_{n+1}^{y} - \sigma_{n-1}^{y} \sigma_{n}^{z} \sigma_{n+1}^{x})
     \Big] -\frac{h}{2} \sum_{n=1}^{N} \sigma_{n}^{z}, \quad
\label{eq:ham}
\end{eqnarray}
where $J$ represents the two-body interaction strength between the
nearest-neighbour spins in XX and YY directions independently, $\alpha$
is the three-body interaction strength between next-next-nearest
neighbour spins, $\gamma$ is the strength of the anisotropy between XX
and YY two-body interactions, $h$ is the external applied magnetic field
in Z-direction across the system, $N$ is the total number of spins arranged in a
one-dimensional array such that the periodic boundary condition, i.e.,
$\sigma_{N+1} = \sigma_1$, is satisfied, and $\sigma_j^a (a = x, y, z)$
are the Pauli spin matrices at the site $j$ of the quantum spin system.

Recently, the shareability of quantum correlations among the subsystems
of this multiparty quantum spin system (with spin chain of length
$N=10$), without the anisotropy constant, has been characterized with
respect to its system parameters, by using both
entanglement-separability and information-theoretic kinds of quantum
correlation measures~\cite{kiran2023shareability}. In the present work,
we consider single channel and various multi channels formed by the
subsystems or the whole of this multiparty quantum spin system with
finite spin chain ($N=10$) for characterizing the multiport quantum
dense coding capacity.

\section{Quantum Dense Coding}\label{sec:qdc}

Quantum dense coding~\cite{bennett1992communication} is a communication
protocol, which uses an entangled state as a quantum channel to send
classical information beyond the capacity of an analogous classical
channel. Let Alice ($A$) and Bob ($B$) share the composite two-party
quantum state $\rho_{AB}$ that belongs to the Hilbert space $H_{A}
\otimes H_{B}$. Alice has to send a classical message $i$, which occurs
with probability $p_i$ at her location to Bob, who is separated from
Alice by a long distance. She encodes this classical message in a
unitary operator $U_i$, which is uniquely identified with the message
$i$, and applies this unitary operator on her part of the composite
system $\rho_{AB}$ to obtain $\rho_{AB}^i = (U_i \otimes \unit)
\rho_{AB} (U_i^{\dag} \otimes \unit)$, where $\unit$ is the identity
operator acting on the subsystem $\rho_B$. After this encoding, she sends
her part of $\rho_{AB}^i$ through a noiseless quantum channel to
Bob. Therefore, Bob has in his possession a two-party ensemble
$\qty{p_i, \rho_{AB}^i}$, and his objective is to acquire maximum
information about the message $i$ by conducting quantum mechanically
allowed measurements on this two-party ensemble. This involves an
optimization procedure which gives a quantifiable form called
quantum dense coding capacity for the state $\rho_{AB}$. The quantum
dense coding capacity for any arbitrary two-partite state $\rho_{AB}$ is
defined as~\cite{bose2000mixed} 
\begin{equation} 
C(\rho_{AB}) = \log_2 d_{A} + S(\rho_{B}) - S(\rho_{AB}), 
\end{equation} 
where $d_A =\dim H_{A}$ is the dimension of the sender subsystem, $S(\sigma) =
-\Tr[\sigma \log_2 \sigma]$ is the von Neumann entropy of the quantum
state $\sigma$ and $\rho_B = \tr_A(\rho_{AB})$ is the reduced density
matrix obtained by tracing out Alice's subsystem from the two-partite
state $\rho_{AB}$. Here the capacity is measured in the units of ``bits''. 

To make this capacity bounded by unity and independent of the dimensions
of the quantum system on which we are working, we divide the
quantum dense coding capacity of the two-partite state by its maximum
quantum mechanically achievable capacity, viz., $\log_2 d_A + \log_2
d_B$. Thus, the {\em normalized single channel quantum dense coding capacity} is given by 
\begin{equation}
	C (\rho_{AB}) = \frac{\log_2 d_{A} + S(\rho_{B}) - S(\rho_{AB})}{\log_2
		d_{A} + \log_2 d_{B}}.
	\label{eq:C_S}
\end{equation}

We now extend this definition of quantum dense coding capacity to
multiport scenario by considering the $N$-party quantum state
$\rho_{A_i B_j}$, with $A_i$'s are its first $i$ particles identified as
the senders of information and $B_j$'s are its latter $j$ particles
identified as the receivers of information. Depending on the number of
$A_i$'s and $B_j$'s, various combinations of multiport quantum dense
coding channels can be generated. For example, multiple senders and a
single receiver ($\rho_{A_i B}$), a
single sender and multiple receivers ($\rho_{A B_j}$), and multiple senders and multiple
receivers ($\rho_{A_i B_j}$). Though several multiport
channels are possible, for the purpose of implementation of multiport quantum
dense coding, we here consider only two cases: multiple senders and
a single receiver, and a single sender and multiple receivers. For
the case of multiple senders and a single receiver, let
$A_1,A_2,\ldots,A_{N-1}$ be the senders and $B$ is a receiver. The
normalized multiport dense coding capacity for this case is the generalization of
Eq.~(\ref{eq:C_S}), applied to $N$-party quantum system
$\rho_{A_1,\ldots,A_{N-1},B}$, and is given
by~\cite{bruss2004distributed, bruss2006dense}
\begin{widetext}
\begin{equation}
  C (\rho_{A_1,\ldots,A_{N-1},B}) = \frac{\log_2 d_{A_1} + \cdots +\log_2 d_{A_{N-1}} +
  S(\rho_{B}) - S(\rho_{A_1,\ldots,A_{N-1},B})}{\log_2 d_{A_1} + \cdots + \log_2 d_{A_{N-1}}  +
  \log_2 d_{B}}.
  \label{eq:C_W}
\end{equation}
\end{widetext}
Here, $d_{A_i}$ is the dimension of the individual sender subsystem
$\rho_{A_i}$ at the $i^{\rm th}$ site of the system, $d_B$ is the
dimension of the single receiver subsystem $\rho_B$, and $\rho_{B} =
\tr_{A_1,\ldots,A_{N-1}}(\rho_{A_1,\ldots,A_{N-1},B})$ is the reduced
density matrix of the receiver subsystem.

For the case of multiport quantum dense coding with a single sender and
multiple receivers (here, $A$ is the sender and $B_1,
B_2,\ldots,B_{N-1}$ are receivers) in a given multiparty quantum system
$\rho_{A, B_1,\ldots,B_{N-1}}$, the exclusion principle of quantum dense
coding comes into action~\cite{prabhu2013exclusion}. The exclusion principle restricts the quantum advantage in this multiport dense coding protocol to dense
coding protocol with only one chosen single channel made up of two of
its parties. All other possible single
channels obtained from the subsystems of this multiparty system are allowed to have up to classical capacity for
transmission of the given classical information. Therefore, the
exclusion principle allows us to define a modified multiport dense
coding capacity for the case of a single sender and multiple receivers
as
\begin{equation}
  C(\rho_{A, B_1,\ldots,B_{N-1}}) = \max \qty{C(\rho_{AB_j})\,|\, j=1,2,\ldots,N-1},
  \label{eq:C_W_adv}
\end{equation}
where $C(\rho_{AB_j})$ is the single channel quantum dense coding
capacity, defined in Eq.~(\ref{eq:C_S}). Here, the single channel is
formed by the subsystem $\rho_{AB_j}$, where $A$ is the first particle
and is a sender, and $B_j$ is any one of the latter
particles which acts as a receiver.

\begin{figure*}[htpb]
\centering
  \subfloat[$\gamma=0$, $h=0.4$]{
  \includegraphics[width=0.25\linewidth]{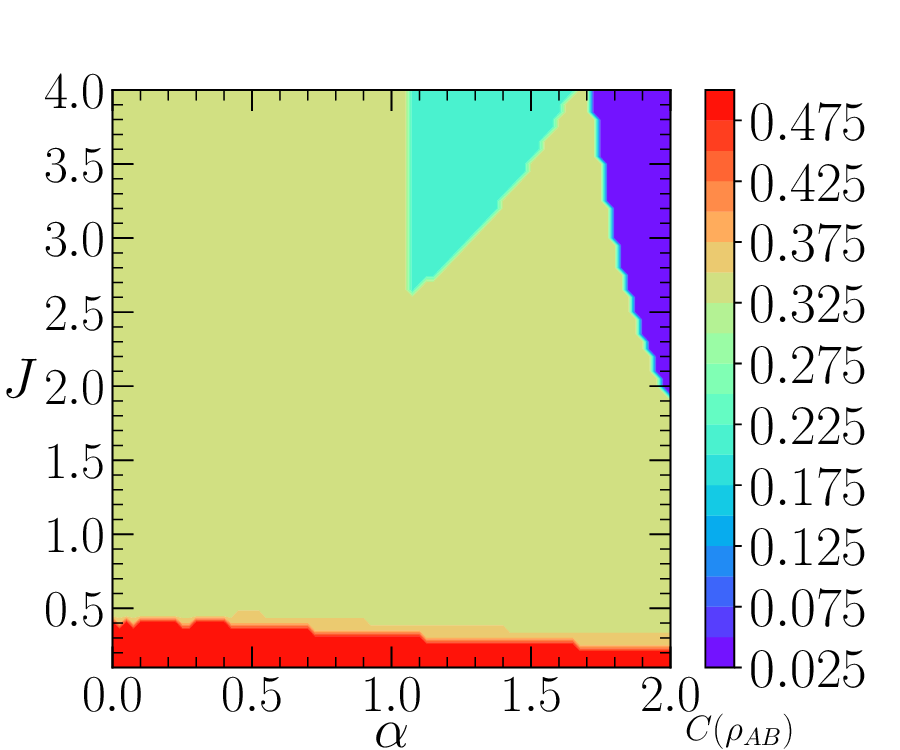}
  }
  \subfloat[$\gamma=0$, $h=0.9$]{
  \includegraphics[width=0.25\linewidth]{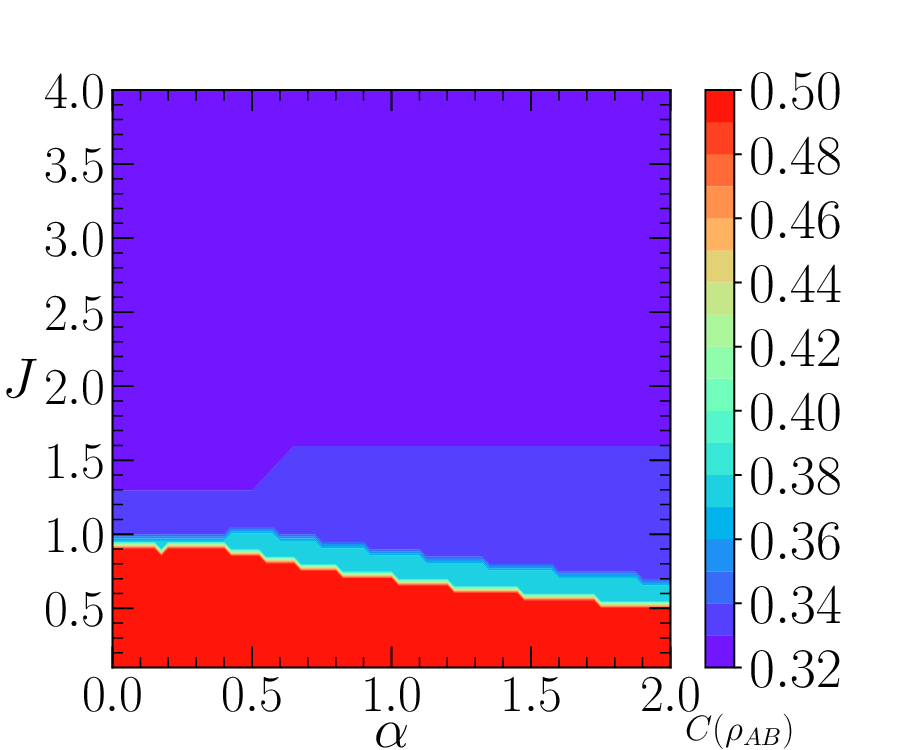}
  }
  \subfloat[$\gamma=0.7$, $h=0.4$]{
  \includegraphics[width=0.25\linewidth]{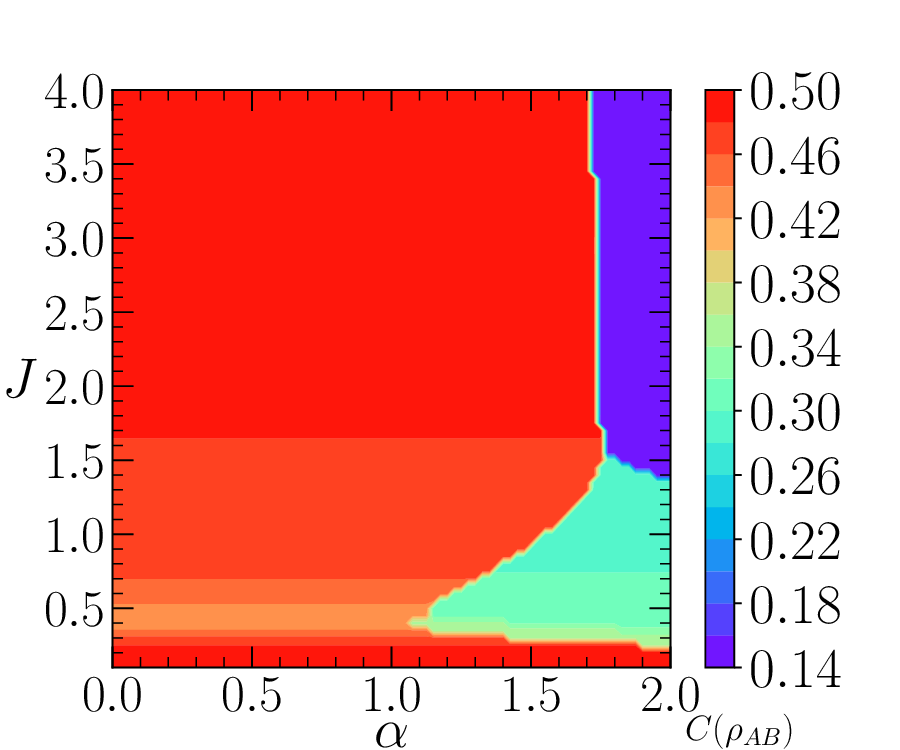}
  }
  \subfloat[$\gamma=0.7$, $h=0.9$]{
  \includegraphics[width=0.25\linewidth]{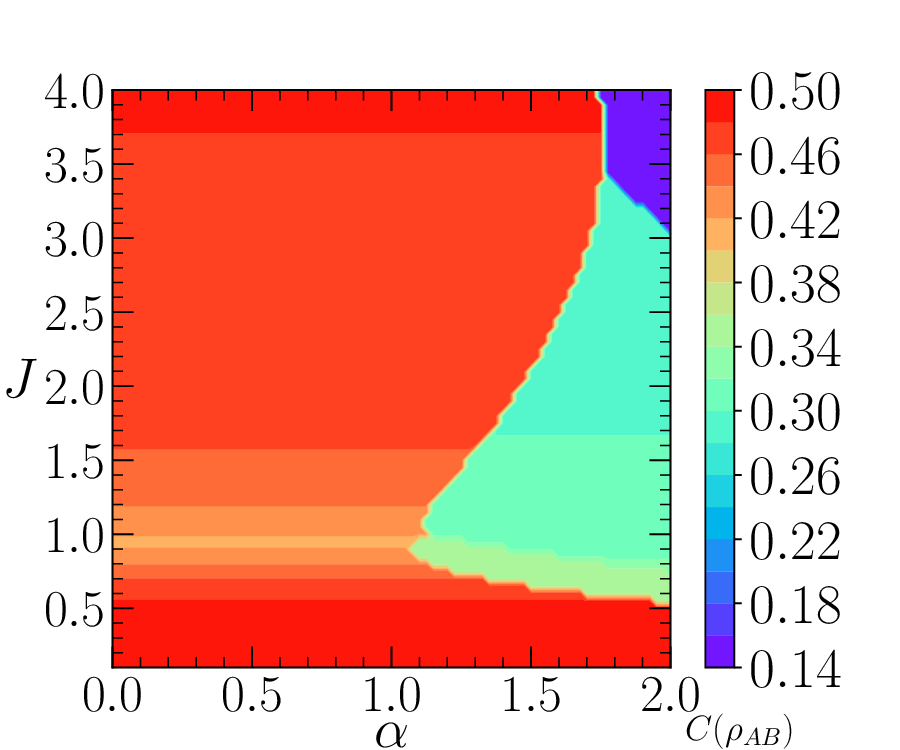}
  }
\caption{(Color online) The quantum dense coding capacity,
$C(\rho_{AB})$, of the single channel formed by the first two spins of
Eq.~(\ref{eq:ham}) with spin chain length of $N=10$ as a function of two-body interaction $J$ and
three-body interaction $\alpha$ for various choices of anisotropy
constant $\gamma$ and external magnetic field strength $h$. Here, (a)
$\gamma=0$, $h=0.4$, (b) $\gamma=0$, $h=0.9$, (c) $\gamma=0.7$, $h=0.4$
and (d) $\gamma=0.7$, $h=0.9$. The single channel quantum dense coding
capacities are always less than or equal to the classical capacity,
irrespective of the ranges of all the system parameters.}
\label{fig:dcc2_N10}
\end{figure*}

\section{Quantum Dense Coding in Multiparty Quantum Spin
System}\label{sec:dcc_isotropic}

The ground state of the anisotropic multiparty quantum spin system with two- and
three-body interactions, for which the Hamiltonian is given in
Eq.~(\ref{eq:ham}), is considered for the creation of various multiport
quantum channels for which we characterize the multiport quantum dense
coding capacity with respect to the system parameters, viz., two-body
interaction $J$, three-body interaction $\alpha$, anisotropy constant
$\gamma$ and external applied magnetic field $h$. The ground state of
the quantum spin system in Eq.~(\ref{eq:ham}) considered here for the chain length of
$N=10$ is obtained by exact diagonalization through numerical
techniques. 

The multiport quantum dense coding protocols can be implemented using
various channels, which are formed by the subsystems obtained from the
marginal density matrices of the ground state of the quantum spin
system or on the multiport channels obtained by
considering the whole system. We now characterize the quantum dense
coding capacity of these various channels obtained from the
anisotropic multiparty quantum spin system with respect to its system
parameters.

\subsection{Single Channel Quantum Dense Coding} 

Let us now consider the single channel formed by the marginal density
matrix $\rho_{AB}$ of the first two spins, which is obtained by tracing
out the latter ($N-2$) spins from the ground state of the $N$-party
quantum spin system. As it is the simplest case of $i=j=1$, we have
ignored
the indices for these parties. Let the first particle $\rho_A$ be the
sender and the second particle $\rho_B$ be the receiver of information
in this single channel quantum dense coding protocol.  We now
characterize the single channel quantum dense coding capacity
$C(\rho_{AB})$ of this two-party system using Eq.~(\ref{eq:C_S}) with
respect to the system parameters of Eq.~(\ref{eq:ham}). The variation of
$C(\rho_{AB})$ with respect to two-body interaction $J$ and three-body
interaction $\alpha$, for various combinations of anisotropy constant
$\gamma$ and the external applied magnetic field $h$ is plotted in
Fig.~\ref{fig:dcc2_N10}.

The choice of plotting single channel quantum dense coding capacity with
the first two nearest-neighbouring spins is for the reason that the other combination of single channels
made up of further distant spins like next-nearest-neighbour,
next-next-nearest-neighbour, etc., obtained by the respective marginal
density matrices from the anisotropic multiparty quantum spin system
under consideration, will possess less entanglement and hence lesser
quantum dense coding capacity than its nearest-neighbour spins~\cite{sachdev2011quantum,
kiran2023shareability}.

We observe that the nearest-neighbour two-qubit system $\rho_{AB}$
obtained from Eq.~(\ref{eq:ham}), which acts as a single channel for
quantum dense coding, will be a negative under partial transpose (NPT)
for the entire range of system parameters. Such class of states that
are mixed and NPT are shown to be useless for quantum dense coding
protocol~\cite{nielsen2000quantum, bruss2004distributed}.  Hence, from
Fig.~\ref{fig:dcc2_N10}, we confirm that $C(\rho_{AB})$ will always
remain less than the classical capacity (0.5) and hence does not
possess any quantum advantage for all ranges of system parameters.
Moreover, another reason for us to shun the studies involving the single
channel quantum dense coding protocol in such multiparty systems is the
concept of {\em receiver monogamy}~\cite{prabhu2013exclusion}, which
states that ``the sum of the single channel quantum dense coding
capacities is always less than the multiport quantum dense coding
capacity with ($N-1$) senders and a single receiver in an $N$-partite
state''. As per receiver monogamy, the sum of all single channel quantum
dense coding capacities, with $B$ being a common receiver, which are
obtained by tracing out remaining ($N-2$) particles at each instance,
including the highest dense coding capacity contained in a single
channel formed by nearest-neighbour two-spin systems, will be less than
the multiport quantum dense coding capacity
$C(\rho_{A_1,\ldots,A_{N-1},B})$.  Therefore, to implement quantum dense
coding protocols in multiparty quantum spin systems like the one in
Eq.~(\ref{eq:ham}), it is advantageous to consider multiport quantum
dense coding protocol instead of single channel quantum dense coding
protocol.

\begin{figure*}[htpb]
\centering
  \subfloat[$\gamma=0$, $h=0.4$]{
  \includegraphics[width=0.25\linewidth]{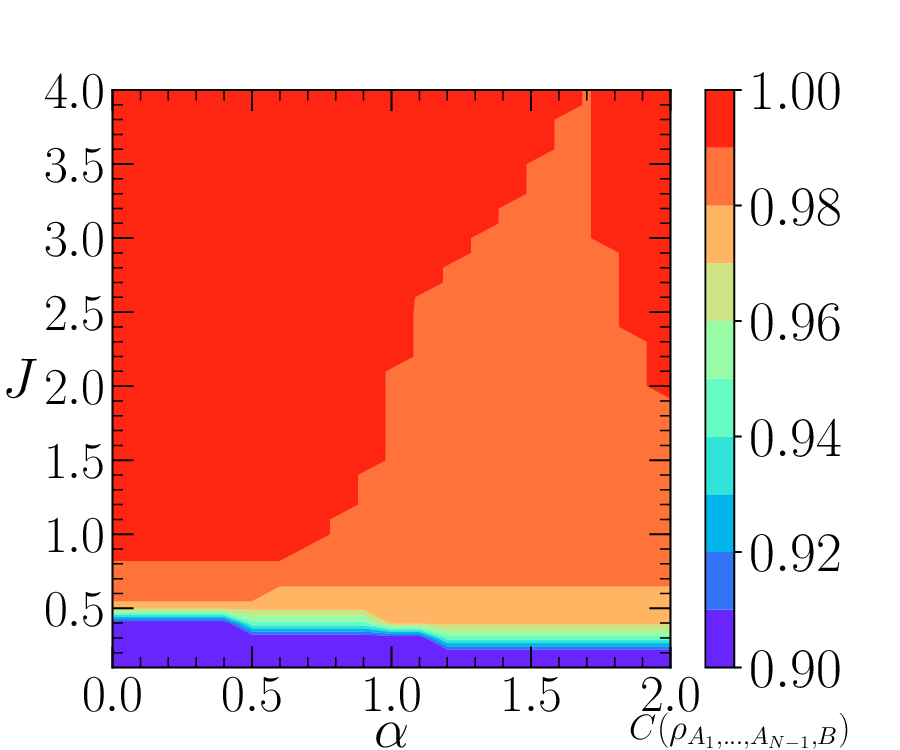}
  }
  \subfloat[$\gamma=0$, $h=0.9$]{
  \includegraphics[width=0.25\linewidth]{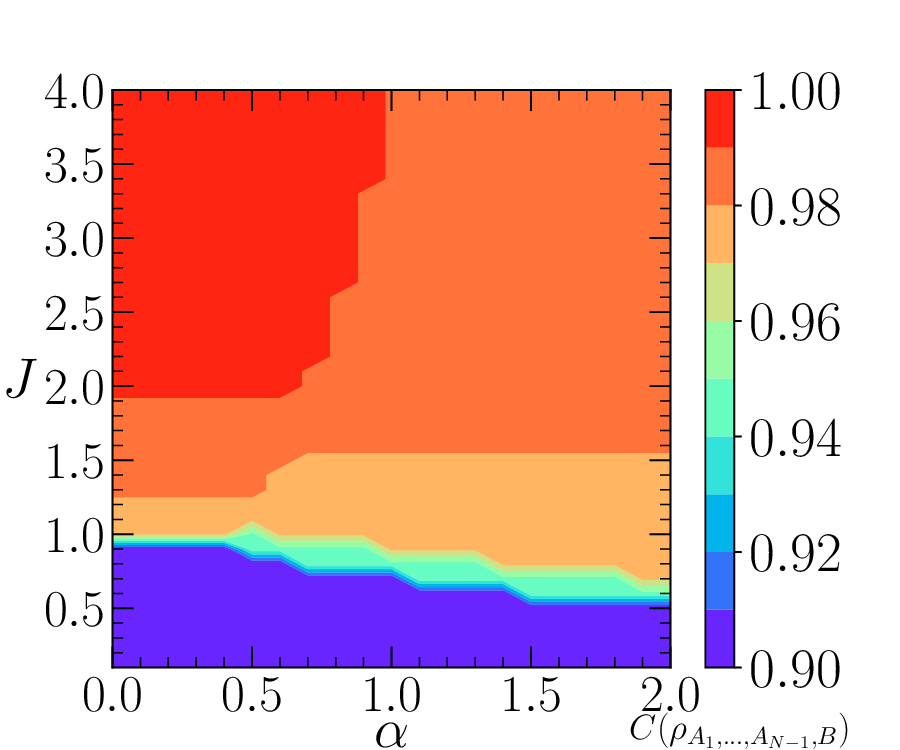}
  }
  \subfloat[$\gamma=0.7$, $h=0.4$]{
  \includegraphics[width=0.25\linewidth]{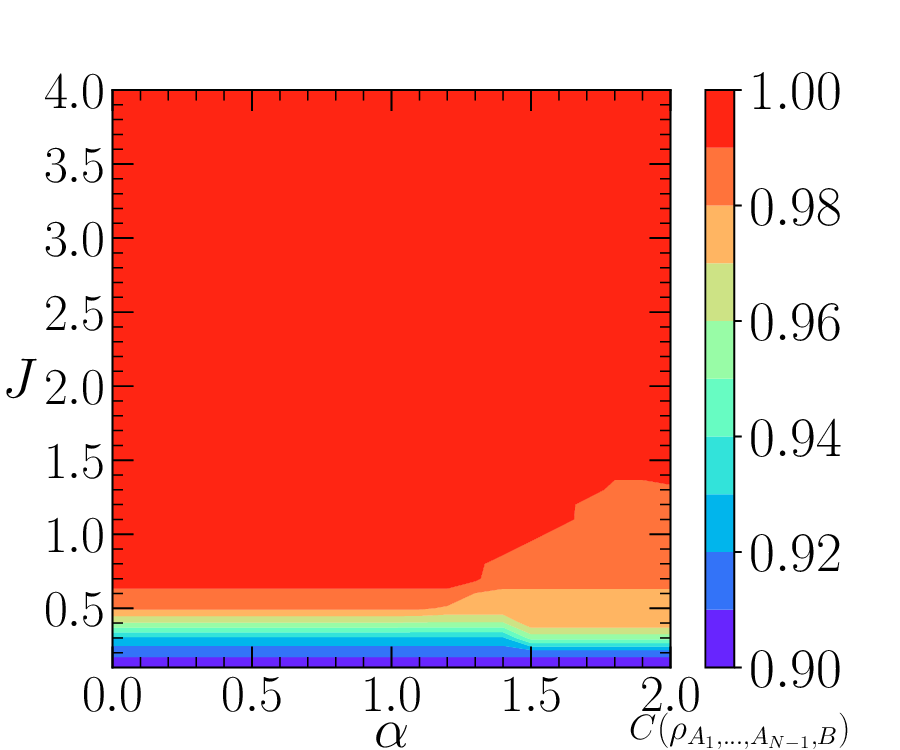}
  }
  \subfloat[$\gamma=0.7$, $h=0.9$]{
  \includegraphics[width=0.25\linewidth]{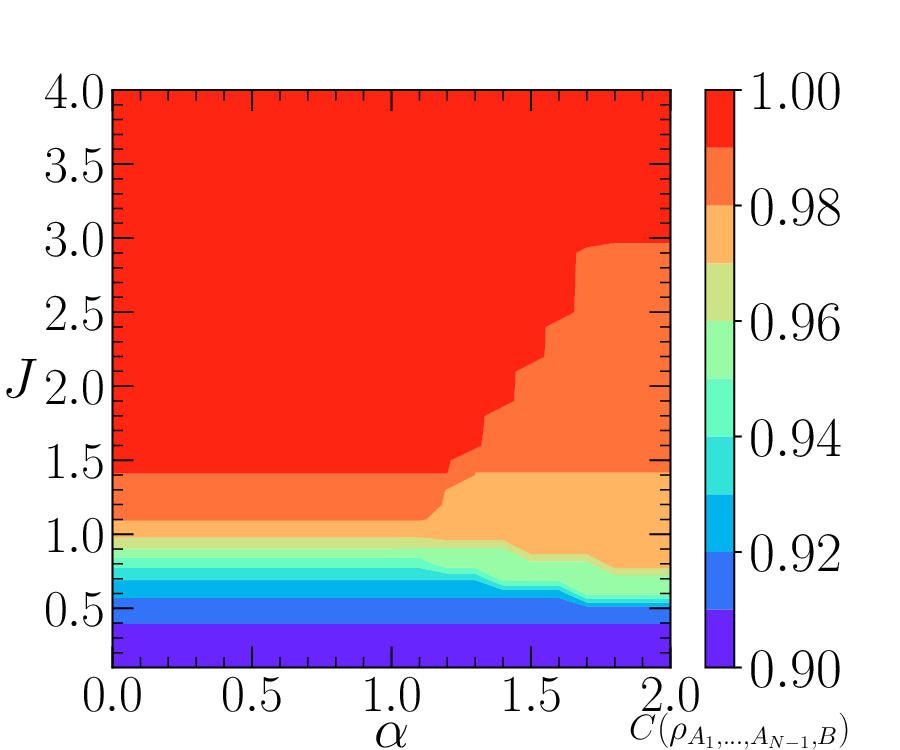}
  }
\caption{(Color online) The quantum dense coding capacities,
$C(\rho_{A_1,\ldots,A_{N - 1}, B})$, of a multiport channel formed by the
($N-1$) senders and a single receiver as a function of two-body interaction
$J$ and three-body interaction $\alpha$ for various choices of
anisotropy constant $\gamma$ and external magnetic field strength $h$. 
Here, (a) $\gamma=0$, $h=0.4$,
(b) $\gamma=0$, $h=0.9$, (c) $\gamma=0.7$, $h=0.4$ and (d) $\gamma=0.7$,
$h=0.9$. $C(\rho_{A_1,\ldots,A_{N-1},B})$ always has the quantum advantage
for all ranges of the system parameters. $\rho_{A_1,\ldots,A_{N - 1},
B}$ is the ground state of Eq.~(\ref{eq:ham}) with $N=10$.}
\label{fig:dccN1_N10}
\end{figure*}

\subsection{Multi Channel Quantum Dense Coding}

In all the possible multiport dense coding scenarios, let us first consider the case of a single
sender and ($N-1$) receivers. The quantum dense coding capacity of such
multiport channels $\rho_{A,B_1,\ldots,B_{N - 1}}$ will be restricted by
the exclusion principle of quantum dense
coding~\cite{prabhu2013exclusion}. The exclusion principle states that,
``in an $N$-partite state, only one of the chosen two-qubit system,
where one qubit is as a sender and the other as receiver, will have a
quantum advantage over all other possible two-qubit subsystems obtained
from that multipartite state''. In
other words, in a multipartite system, among all the possible single
channels formed by its subsystems, there exists only one chosen
quantum channel which exhibits the quantum dense coding, and all other
quantum channels
formed by its other subsystems can have utmost classical capacity for
transmission of classical information from the sender to the receiver.
In the current anisotropic multiparty quantum spin system given in
Eq.~(\ref{eq:ham}), the highest capacity for transmitting classical
information via a single quantum channel is less than the classical capacity.
Due to this observation, along with exclusion principle for dense
coding, we no longer consider the case of one sender and ($N-1$)
receivers multiport quantum dense coding protocol for our studies. This
allows us to conclude that when multiparty quantum spin systems like
Eq.~(\ref{eq:ham}) with marginal two spin states are all NPT, we can
restrict the study of multiport quantum dense coding capacity to only
the case of ($N-1$) senders and a single receiver. All other multiport
quantum dense coding protocols will yield quantum dense coding capacity
which is either equal to or less than that of the classical capacity
($C_{\rm Cl}$) of the same multiport channels.

Let us now consider the other case of multiport quantum dense coding
protocol with ($N-1$) senders and a single
receiver multiport channel used in the multiport quantum dense coding
protocol for which the capacity is given by Eq.~(\ref{eq:C_W}). Note
that, for a spin chain of length $N$, Eq.~(\ref{eq:C_W}) will be of the
form
\begin{equation}
  C(\rho_{A_1,\ldots,A_{N - 1},B}) = \frac{N-1 + S(\rho_{B}) -
  S(\rho_{A_1,\ldots,A_{N - 1},B})}{N}.
\end{equation}
Here, we have used the fact that, for spin-$1/2$ particles, $\log_2
d_{A_i} = 1$ and there are ($N-1$) senders. For classical channels, the
above equation will be equal to $(N - 1)/N$ as there is no quantum
advantage ($S(\rho_{B}) -  S(\rho_{A_1,\ldots,A_{N - 1},B})=0$).  When
there is a quantum advantage, i.e., ($0 < S(\rho_{B}) -
S(\rho_{A_1,\ldots,A_{N - 1},B}) \leq 1$), the above equation will vary
from $(N-1)/N$ to $1$. Therefore, for the system under consideration
with spin chain length of $N=10$, when there exists a quantum dense
coding advantage for the channel with ($N-1$) senders and a single
receiver, the multiport quantum dense coding capacity varies from $0.9$
to $1$.  

\begin{figure*}[htpb]
\centering
  \subfloat[$\alpha=0.6$, $\gamma=0$]{
  \includegraphics[width=0.25\linewidth]{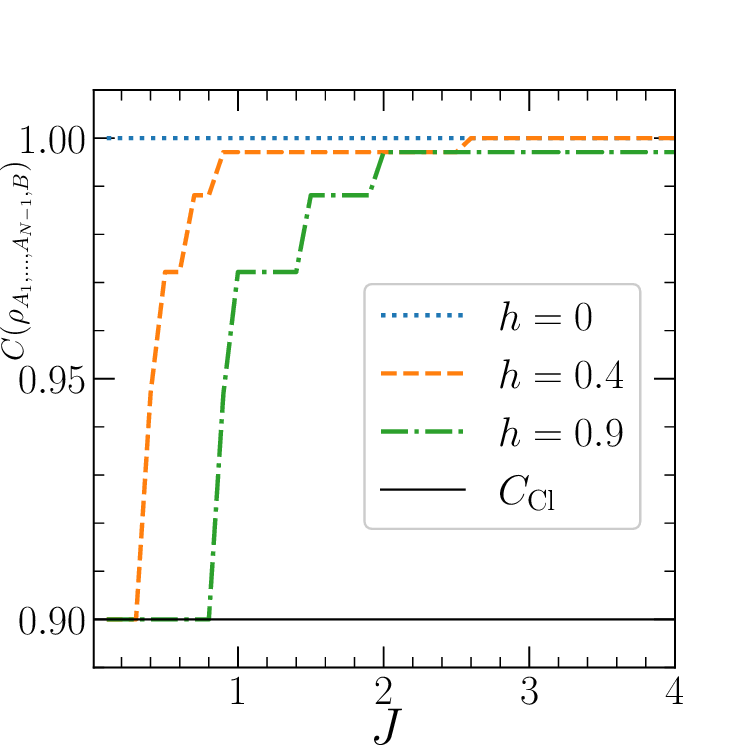}
  }
  \subfloat[$\alpha=0.6$, $\gamma=0.7$]{
  \includegraphics[width=0.25\linewidth]{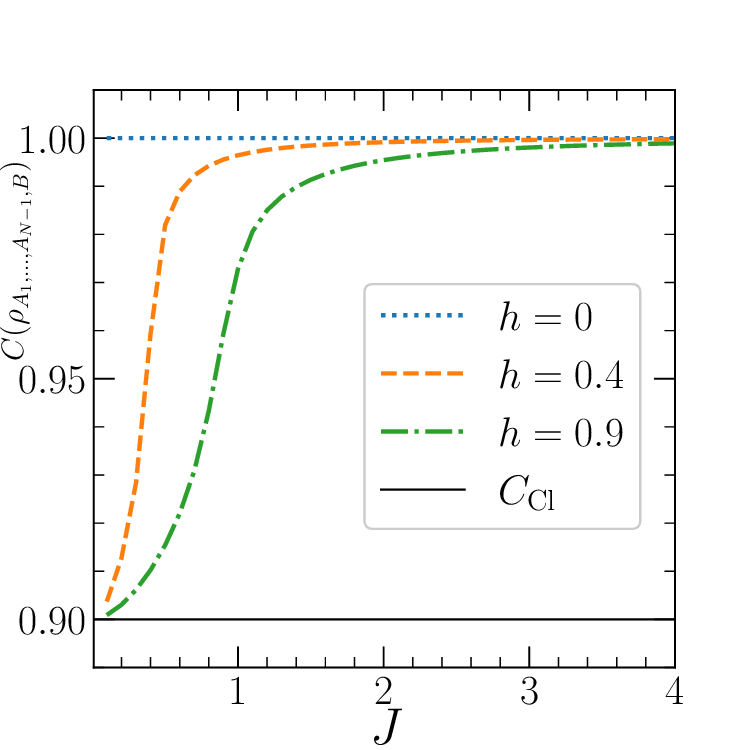}
  }
  \subfloat[$\alpha=1.8$, $\gamma=0$]{
  \includegraphics[width=0.25\linewidth]{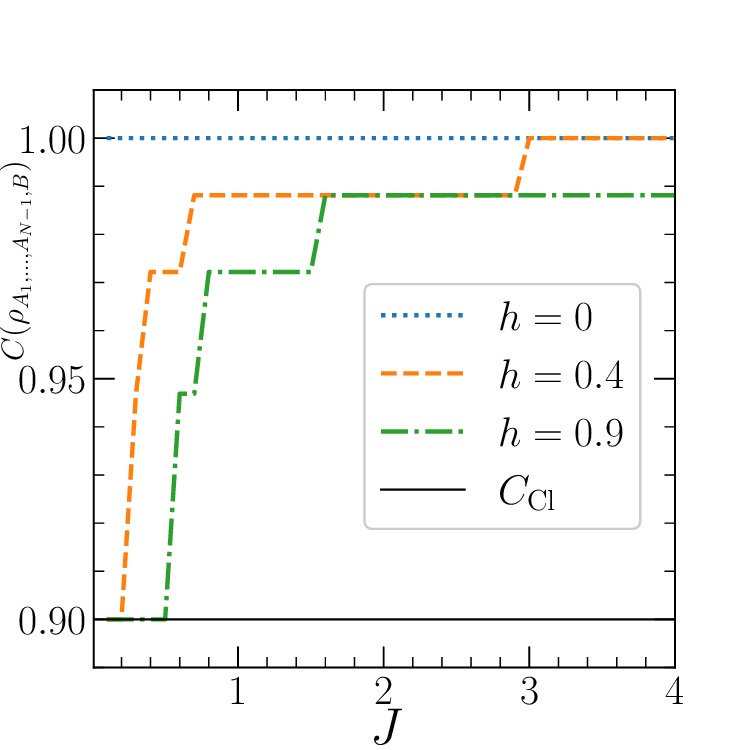}
  }
  \subfloat[$\alpha=1.8$, $\gamma=0.7$]{
  \includegraphics[width=0.25\linewidth]{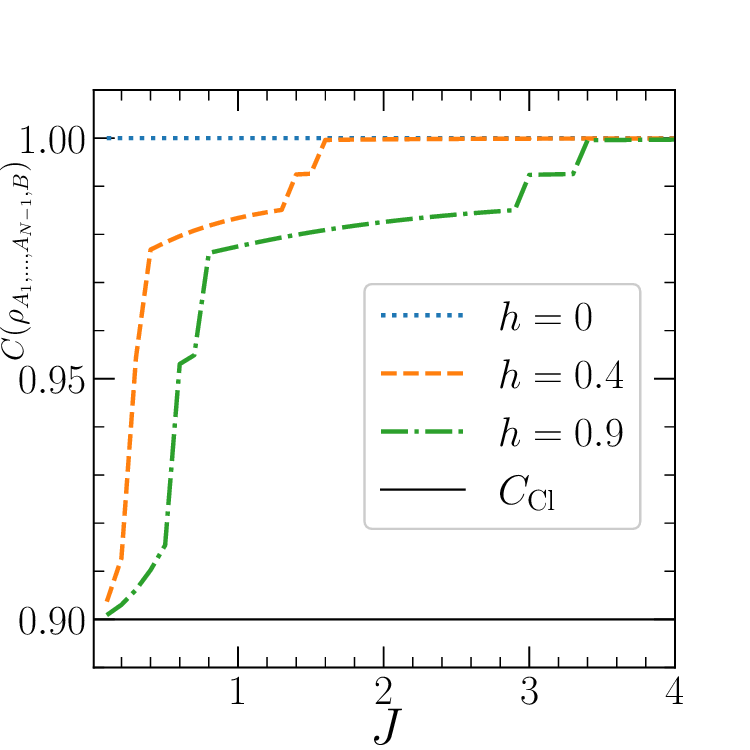}
  }
\caption{(Color online) The quantum dense coding capacities,
$C(\rho_{A_1,\ldots,A_{N - 1}, B})$, of a multiport channel formed by the
($N-1$) senders and a single receiver as a function of two-body
interaction $J$ for various choices of three-body interaction $\alpha$
and anisotropy constant $\gamma$. Here, (a) $\alpha=0.6$, $\gamma=0$,
(b) $\alpha=0.6$, $\gamma=0.7$, (c) $\alpha=1.8$, $\gamma=0$ and (d)
$\alpha=1.8$, $\gamma=0.7$ for external magnetic field strengths of
$h=0,0.4$ and $0.9$. The solid (black) line $C_{\rm Cl}$ represents the
classical capacity of the channel. In the absence of the external
magnetic field, $C(\rho_{A_1,\ldots,A_{N - 1}, B})$ always has the maximum
value of 1 (blue dotted line). For, higher two-body interaction $J$,
$C(\rho_{A_1,\ldots,A_{N - 1}, B})$ approaches maximum value irrespective of
all the other system parameters. $\rho_{A_1,\ldots,A_{N - 1},
B}$ is the ground state of Eq.~(\ref{eq:ham}) with $N=10$.}
\label{fig:dcc_N10}
\end{figure*}

To observe this, we plot $C(\rho_{A_1,\ldots,A_{N - 1}, B})$ with
respect to the variation of two-body interaction $J$ and three-body
interaction $\alpha$, for chosen values of anisotropy constant $\gamma$
and the external applied magnetic field strength $h$ in
Fig.~\ref{fig:dccN1_N10}. The choice of $\gamma$ and $h$ are due to the
more pronounced features of $C(\rho_{A_1,\ldots,A_{N - 1},B})$ at these
values over the others with less variations with respect to $J$ and
$\alpha$. From Fig.~\ref{fig:dccN1_N10}, it is clear that
$C(\rho_{A_1,\ldots,A_{N -
1},B})$ always beats the classical capacity in the multiparty spin
  system under consideration for the entire ranges of system parameters
  of Eq.~(\ref{eq:ham}). Since, the behaviour of
  $C(\rho_{A_1,\ldots,A_{N - 1},B})$ is more involved with respect to
  the variation in $J$ than the variation in $\alpha$, we probe
  $C(\rho_{A_1,\ldots,A_{N - 1},B})$ with respect to $J$ for specific
  values of $\alpha$ and $\gamma$ in Fig.~\ref{fig:dcc_N10}.  These
  choices of $\alpha$ and $\gamma$ are due to better exhibition of
  characteristics of $C(\rho_{A_1,\ldots,A_{N - 1},B})$ in the plots
  with respect to the system parameters. For other choices of $\alpha$
  and $\gamma$, the observations
  of $C(\rho_{A_1,\ldots,A_{N - 1},B})$ remain similar. From Fig.~\ref{fig:dcc_N10}, the following conclusions can be drawn on the capacity of
  multiport quantum dense coding protocol with ($N-1$) senders and a
  single receiver: (i) When an external applied magnetic field is absent
  ($h=0$), $C(\rho_{A_1,\ldots,A_{N - 1},B})$ always remains maximum
  value of 1 and it does not depend on the value of $J$, $\alpha$ and
  $\gamma$. (ii) In the presence of external magnetic field,
  $C(\rho_{A_1,\ldots,A_{N -
1},B})$ varies from classical capacity to maximum with respect to
  increase in $J$, and this observation remains true for all $\alpha$
  and $\gamma$. (iii) $C(\rho_{A_1,\ldots,A_{N - 1},B})$ for lower $h$
  will always be greater than the same for higher $h$ with respect to
  $J$, and this behaviour is independent of $\alpha$ and $\gamma$. (iv)
  $C(\rho_{A_1,\ldots,A_{N -
1},B})$ reaches its maximum value quickly with respect to $J$ for low
  values of $h$. (v) As the strength of $\alpha$ is increased,
  $C(\rho_{A_1,\ldots,A_{N - 1},B})$ reaches maximum at higher values of
  $J$. To consolidate, the multiport quantum dense coding capacity
  with ($N-1$) senders and a single receiver will attain maximum value
  either when no external magnetic field strength $h$ is applied across
  the system or when two-body interaction $J$ is higher for any value of
  other system parameters contained in Eq.~(\ref{eq:ham}).

\subsection{Magnetic Field Averaged Dense Coding Capacity}

The characteristics of multiport quantum dense coding capacity
consisting of multiport quantum channels with ($N-1$) senders and a single
receiver for the given system is observed to be similar for any strength of the
external applied magnetic field $h$, as captured in
Fig.~\ref{fig:dcc_N10}. These characteristics remain same for
the variation of other system parameters of Eq.~(\ref{eq:ham}). In other words,
$C(\rho_{A_1,\ldots,A_{N-1},B})$ remains equal to classical capacity for
low $J$ and reaches maximum for high $J$, and at intermediate values of
$J$ there is a steady increase in $C(\rho_{A_1,\ldots,A_{N - 1},B})$ for all
strengths of $h$. Therefore, we take an average of
$C(\rho_{A_1,\ldots,A_{N - 1},B})$ for all the values of external applied magnetic
field and define a new quantity called {\em magnetic field averaged
dense coding capacity}, which is given by
\begin{equation}
  \langle C_h (\rho_{A_1,\ldots,A_{N-1},B}) \rangle = \frac{1}{m}\sum_{i}^{m} C(\rho_{A_1,\ldots,,A_{N-1},B} (h_i)),
  \label{eq:C_avg}
\end{equation}
where $C(\rho_{A_1,\ldots,A_{N-1},B} (h_i))$ is the individual multiport
quantum dense coding capacity with ($N-1$) senders and a single
receiver, computed for the ground state of the system under
consideration when an external magnetic field of strength $h_i$ is
applied across the whole system given in Eq.~(\ref{eq:ham}). $\langle
C_h (\rho_{A_1,\ldots,A_{N-1},B}) \rangle$, henceforth denoted as
$\chavg$, is the multiport quantum dense coding capacity averaged over
the entire range of magnetic field strengths within the range
$0<h_i\leq1$.  To calculate $\chavg$, we obtain the ground state of
Eq.~(\ref{eq:ham}), evaluate $C (\rho_{A_1,\ldots,A_{N-1},B} (h_i))$
using Eq.~(\ref{eq:C_W}) for $100$ equidistant values of $h_i$ chosen
between $0<h_i\leq1$ and then calculate their average. This averaged
quantity will help us to comprehensively capture the influence of the
entire range of magnetic field strengths on the characteristics of
multiport quantum dense coding capacity with ($N-1$) senders and a
single receiver with respect to the system parameters of
Eq.~(\ref{eq:ham}). 

\begin{figure}[htpb]
\centering
  \subfloat[$\gamma=0$]{
  \includegraphics[width=0.5\linewidth]{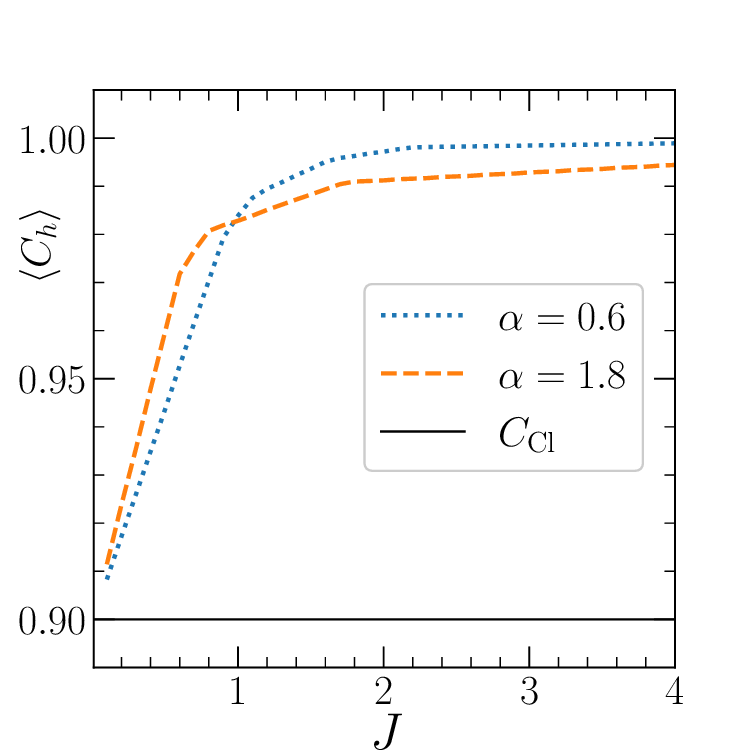}
  }
  \subfloat[$\gamma=0.7$]{
  \includegraphics[width=0.5\linewidth]{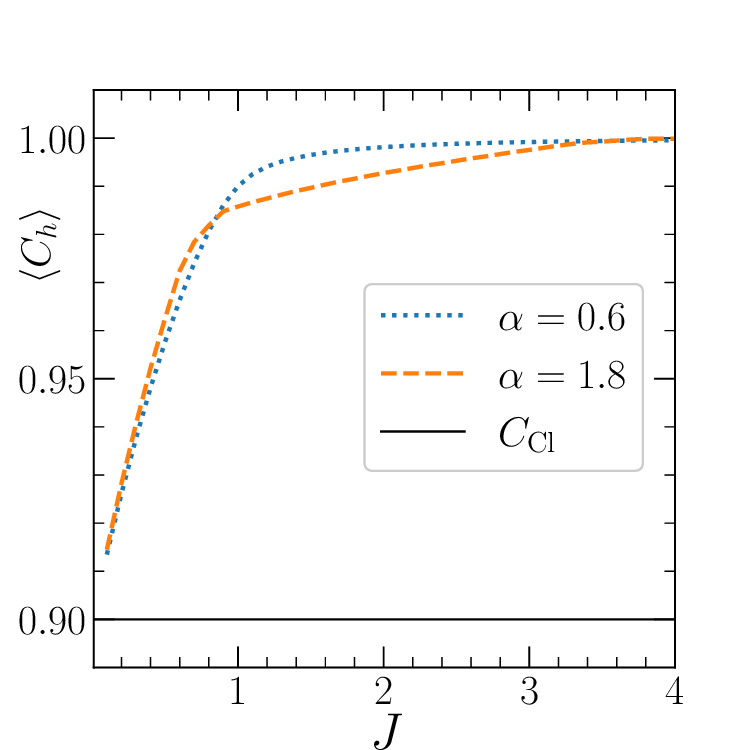}
  }
\caption{(Color online) Magnetic field averaged dense coding capacities,
$\chavg$, for the anisotropic multiparty quantum spin system given in
Eq.~(\ref{eq:ham}), when ($N-1$) spins are the senders and the last spin
being the receiver of information, as a function of two-body interaction
$J$ for various choices of anisotropy constant $\gamma$. Here, (a)
$\gamma=0$ and (b) $\gamma=0.7$, for different values of three-body
interaction $\alpha$. The solid (black) line $C_{\rm Cl}$ represents the
classical capacity of the channel. $\chavg$ will be maximum, when
$0<J<1$: high $\alpha$ and high $\gamma$ or when $1<J<4$: low $\alpha$
and high $\gamma$. Here, $N=10$ is considered as the spin chain length.}
\label{fig:dcc_averaged_N10}
\end{figure}

\begin{figure}[htpb]
\centering
  \includegraphics[width=0.8\linewidth]{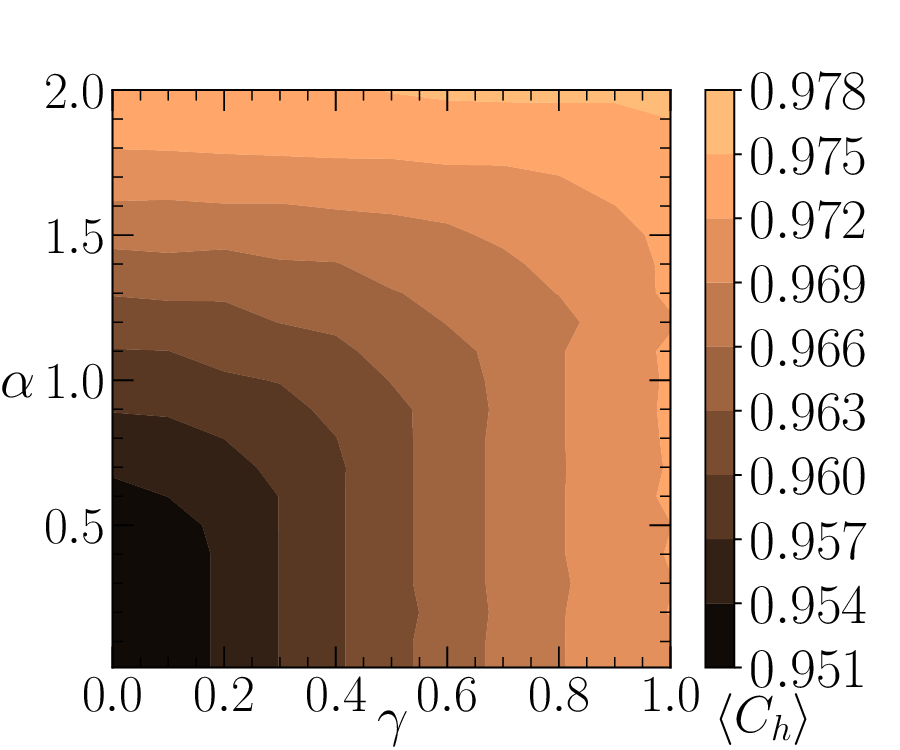}
\caption{(Color online) Magnetic field averaged dense coding capacity,
$\chavg$, as a function of anisotropy constant $\gamma$ and three-body
interaction $\alpha$, for the specific value of two-body interaction
$J=0.6$. In the low $J$ region, $\chavg$ will be lower when both
$\alpha$ and $\gamma$ are lower, and $\chavg$ increases if either of
$\alpha$ or $\gamma$ is increased. Here, to evaluate $\chavg$, we have
chosen the spin chain length of $N=10$ in Eq.~(\ref{eq:ham}).}
\label{fig:contour_dcc_average_N10_J6}
\end{figure}

In Fig.~\ref{fig:dcc_averaged_N10}, the $\chavg$ for our system with
$N=10$ is plotted by continuously varying two-body interaction $J$ and
for chosen values of three-body interaction $\alpha=0.6$ and $1.8$.
These choices of $\alpha$ are due to better exhibition of
characteristics of $\chavg$ in the plots with respect to the system
parameters. For other choices of $\alpha$, at low and high values, the
observations of $\chavg$ remain similar. The following features are
observed with respect to $\chavg$ in the system under consideration with
two- and three-body interaction for which the Hamiltonian is given in
Eq.~(\ref{eq:ham}): (i) For low $J$ values, $\chavg$ will be close to
classical capacity and tends to maximum value as $J$ is increased. (ii)
For lower $J$ values, $\chavg$ for three-body interaction $\alpha=1.8$
will be greater than the same for $\alpha=0.6$.  Around $J=1$, this
feature reverses with $\chavg$ for $\alpha=0.6$ will be greater than the
same for $\alpha=1.8$. (iii) As anisotropy constant $\gamma$ increases,
at low $J$ values, i.e., $J \leq 1$, the divergence between $\chavg$ for
$\alpha=0.6$ and $1.8$ will reduce. (iv) When $J < 1$, the variation of
$\chavg$ irrespective of $\alpha$ and $\gamma$ is steeper. However in
the $J>1$ region, the $\chavg$ reaches its maximum value slowly. (v) There is a maximum
divergence of $\chavg$ around $J=0.6$. Therefore we now plot $\chavg$ with respect to
$\alpha$ and $\gamma$ at $J=0.6$ in
Fig.~\ref{fig:contour_dcc_average_N10_J6} to capture the maximum
divergence in $\chavg$ with respect to variation in $\alpha$ and
$\gamma$. From this figure, it is clear that at low
two-body interaction strengths, the characteristics of $\chavg$ is
involved for low $\alpha$ (and the entire range of $\gamma$) and low
$\gamma$ (and the entire range of $\alpha$). The $\chavg$ will be lower
at both $\alpha$ and $\gamma$ being lower valued, and $\chavg$ will be
higher if either of $\alpha$ or $\gamma$ is larger. 

To consolidate, for the anisotropic multiparty quantum spin system with
two- and three-body interactions for which Hamiltonian is given in
Eq.~(\ref{eq:ham}), the magnetic field averaged dense coding capacity
will be maximum for the choice of low strength of two-body interaction and high
values of three-body interaction and anisotropy (low $J$, high $\alpha$ and
high $\gamma$), or high strength of two-body interaction and low
strength of three-body interaction
and high values of anisotropy (high $J$, low $\alpha$ and high $\gamma$).

\section{Conclusion}\label{sec:conclusion}

Many-body quantum spin systems form reliable physical resources to
implement quantum communication protocols involving multiple senders and
receivers. We considered an anisotropic multiparty quantum spin system
with two- and three-body interactions in the presence of an external
applied magnetic field to characterize the quantum dense coding in
various quantum channels that could be generated by this system.  In our
initial exploration, we focused on the single-channel scenario of
quantum dense coding using the subsystems of the multiparty quantum spin
system considered here. It is well known that the single channels
derived from the subsystems of nearest-neighbouring spins exhibit the
highest capacities for transmitting information from a single sender to
a single receiver when compared to other possible single channels
produced in any multiparty quantum spin systems. In our multiparty
quantum spin system, single channel quantum dense coding capacity
remained less than that of the classical capacity for all ranges of the
system parameters. Further, the ground state of the anisotropic
multiparty quantum spin system considered here, as a whole is employed
for the multiport quantum dense coding protocol. We found that, owing to
the fact that the value of single channel quantum dense coding being
less than classical capacity and exclusion principle of quantum dense
coding, the case of the single sender and multiple receivers is
equivalent to the single channel case and thus lacks any quantum
advantage. Therefore, we established that in the given multiparty
quantum spin system, only the multiport quantum dense coding protocol with
($N-1$) senders and a single receiver exhibits the quantum advantage
over all the other scenarios of multiport quantum dense coding
protocols.

We then characterized the multiport quantum dense coding capacities with
($N-1$) senders and a single receiver, which utilizes the whole ground
state of the system under consideration which is given in
Eq.~(\ref{eq:ham}), with respect to its system parameters. Our findings
indicate that the multiport quantum dense coding capacity for this
configuration reaches its maximum value of 1 either in the absence of an
external magnetic field strength $h$ or when the two-body interaction
strength $J$ is higher, irrespective of the values of the other system
parameters. Notably, we observed that this characteristics of the
multiport quantum dense coding capacity remained consistent across all
external magnetic field strengths $h$ for the entire range of two-body
interaction $J$, and for any chosen values of other system parameters.
To comprehensively assess the impact of the entire range of external
applied magnetic field strengths on the multiport quantum dense coding
capacity with respect to system parameters, we introduced a quantity
called {\em magnetic field averaged dense coding capacity} ($\chavg$).
Our investigation of this averaged quantity for the spin chain length of
$N=10$ revealed that, to enhance multiport quantum dense coding capacity
in the low two-body interaction region ($J<1$), choosing high values of
three-body interaction and high values of anisotropy is advantageous.
Whereas in the high two-body interaction region ($J>1$), higher
multiport quantum dense coding capacity can be achieved by considering
low values of three-body interaction and high values of anisotropy. 

In conclusion, in the implementation of multiport quantum dense coding
protocol using multiparty quantum spin systems, we identified that only
the multiport channel formed by considering the whole ground state and
by choosing ($N-1$) senders and a single receiver within the system has
quantum advantage over other quantum dense coding scenarios. We also
identified the ranges of system parameters required to achieve maximum
capacity of the multiport channel. Our research work hopes to illuminate
the path for further explorations into other multiport quantum
communication protocols using multiparty quantum spin systems and
further strengthen the connection between many-body physics and quantum
information.

\begin{acknowledgments}
RP and HSH acknowledge financial support from the Science and
Engineering Research Board (SERB), Government of India, under the
project grants CRG/2018/004811 and CRG/2021/008795. We acknowlegde the
computations performed at \textit{AnantGanak}, the High Performance
Computing facility at IIT Dharwad.
\end{acknowledgments}

\newpage
\bibliography{reference}
\bibliographystyle{apsrev4-2}
\end{document}